\documentclass[a4paper,twocolumn,11pt,accepted=2020-04-27]
{quantumarticle}
\pdfoutput=1
\usepackage[numbers,compress]{natbib}
\usepackage[utf8]{inputenc}
\usepackage[english]{babel}
\usepackage[T1]{fontenc}

\usepackage{hyperref}

\usepackage{amsmath}
\usepackage{amssymb}
\usepackage{mathtools}
\usepackage{graphicx}
\usepackage{braket}
\usepackage{paralist}
\usepackage{color}
\usepackage{amsthm}
\usepackage{thmtools}
\usepackage{xcolor}
\usepackage{soul}

\usepackage{centernot}
\usepackage{stmaryrd}

\usepackage{tikz}
\usepackage{lipsum}

\theoremstyle{plain}

\newtheorem{proposition}{Proposition}
\newtheorem*{proposition*}{Proposition}

\newtheorem*{theorem*}{Theorem}

\DeclareMathOperator*{\Tr}{Tr}

\DeclareMathOperator*{\Ker}{Ker}
\DeclareMathOperator*{\Pur}{Pur}
\DeclareMathOperator*{\Real}{Re}

\DeclareMathOperator*{\Symm}{Sym}

\newcommand{\bs}{\boldsymbol}

\newcommand{\sz}{\sigma_z}
\newcommand{\ml}[1]{\mathsf{L}_{#1}}
\newcommand{\mr}[1]{\mathsf{R}_{#1}}

\newcommand{\cqf}[4]{\llbracket #1 \rrbracket _{ #2, #3}^{(#4)}}
\newcommand{\cqt}[3]{\llbracket #1 \rrbracket _{ #2}^{(#3)}}
\newcommand{\cq}{\llbracket \rho \rrbracket _{\mathcal M,\mathcal N}^{(\lambda)}}

\makeatletter

\begin{document}

\title{Symmetries and monotones in Markovian quantum dynamics}

\author{Georgios Styliaris}
\email{styliari@usc.edu}
\affiliation{Department of Physics and Astronomy, and Center for Quantum Information Science and Technology, University of Southern California, Los Angeles, California 90089-0484, USA}
\author{Paolo Zanardi}

\affiliation{Department of Physics and Astronomy, and Center for Quantum Information Science and Technology, University of Southern California, Los Angeles, California 90089-0484, USA}

\maketitle

\begin{abstract}

What can one infer about the dynamical evolution of quantum systems just by symmetry considerations? For Markovian dynamics in finite dimensions, we present a simple construction that assigns to each symmetry of the generator a family of scalar functions over quantum states that are monotonic under the time evolution. The aforementioned monotones can be utilized to identify states that are non-reachable from an initial state by the time evolution and include all constraints imposed by conserved quantities, providing a generalization of Noether's theorem for this class of dynamics. As a special case, the generator itself can be considered a symmetry, resulting in non-trivial constraints over the time evolution, even if all conserved quantities trivialize. The construction utilizes tools from quantum information-geometry, mainly the theory of monotone Riemannian metrics. We analyze the prototypical cases of dephasing and Davies generators. 

\end{abstract}

\section{Introduction}

One of the main tasks in the study of non-relativistic quantum dynamical systems is predicting how quantum states and observables evolve over time given some dynamical law, for instance, a Hamiltonian operator and the associated equations of motion.
It is often the case, however, that in practice the trajectory either does not admit an explicit closed form, or even if it does, it can be complicated to draw conclusions to physical questions of interest from it. It is therefore important to have accessible methods and tools that allow one to extract the essential features of the evolution directly from the dynamical laws.

At the heart of such approaches lies the far-reaching idea of symmetry. The most prominent example is perhaps provided by Noether's theorem which, in the context of Lagrangian mechanics, yields a conserved quantity for each differentiable symmetry of the generator~\cite{landau1959classical}. In the quantum realm, the theorem indicates that all moments of an observable that is a symmetry of the time evolution are conserved~\cite{sakurai2017modern}. 

In this paper we consider open quantum systems in finite dimensions evolving under Markovian dynamics~\cite{breuer2002theory}.
%, described by a time-independent generator $\mathcal L$.
We address the question: \textit{Given the generator of the dynamics and an initial state, what constraints can symmetry considerations impose on the set of states that are reachable under the time evolution?}

Open quantum dynamics is, in general, dissipative. As such, the role of conserved quantities can be rather limited. For instance, Markovian dynamics with a unique steady state does not admit any non-trivial conserved quantities~\cite{albert2014symmetries}. For this reason, we approach the problem by instead seeking to utilize symmetries to obtain monotones, i.e., functions of the quantum state that are monotonic (in our case, non-increasing) under the time evolution. Similarly to conserved quantities, monotones can be utilized to exclude state transitions that are impossible under the dynamics, i.e., states that do not belong in the trajectory of a given initial state.

An accessible discussion regarding the connection between symmetries and conserved quantities for Markovian dynamics can be found in the papers by Baumgartner and Narnhofer~\cite{Baumgartner2008analysis}, and by Albert and Jiang~\cite{albert2014symmetries}. Symmetries have also been discussed for the closely related case of iterated quantum channels~\cite{albert2019asymptotics}.
%General considerations regarding the connection between symmetries and conserved quantities for the case of Markovian dynamics have been discussed by Baumgartner and Narnhofer~\cite{Baumgartner2008analysis}, by Albert and Jiang~\cite{albert2014symmetries},
%\cite{baumgartner2008analysis,buca2012note,albert2014symmetries}
%and also for the closely related case of repeated iterations of a quantum channel more recently by Albert~\cite{albert2019asymptotics}.
%, for dissipative spin chains in Ref.~\cite{buca2012note}
%by Baumgartner and Narnhofer in~\cite{Baumgartner2008analysis} and also Albert and Jiang in~\cite{albert2014symmetries}. 
From the viewpoint of quantum resource theories~\cite{chitambar2019quantum}, consequences of symmetries in quantum dynamics have been systematically considered in the theory of reference frames and asymmetry~\cite{bartlett2007reference,marvian2013theory}.  There, one investigates the allowed state transitions under quantum dynamics that respects a specified symmetry\footnote{More specifically, given a group $G$, one investigates the possibility of state transitions under the class of quantum operations that are symmetric with respect to a unitary representation of $G$, namely the operations such that $\left[ \mathcal E , \mathcal U_g \right] = 0$ $\forall g \in G$.}. In particular, Marvian and Spekkens established the fact that, for the case of Hamiltonian dynamics, asymmetry monotones yield conserved quantities that can be independent of the Noether ones~\cite{marvian2014extending}. Asymmetry monotones have also been utilized to put constraints on the evolution of quantum coherences by Lostaglio~et~al.~\cite{lostaglio2017markovian}.  Symmetry considerations have moreover been invoked to study transport in a Markovian model of open spin chains by Bu{\v{c}}a and Prosen~\cite{buca2012note}.

Monotones generalize the concept of a conserved quantity and, as we show in our construction, in fact one can deduce from them the full set of conserved quantities. Our main result consists of a method to assign to each pair of symmetries of the generator a one-parameter family of monotones for the time evolution. As a special case, the generator itself can be considered a symmetry of the dynamics, resulting in non-trivial constraints over the time evolution, even if all conserved quantities trivialize.

The basic idea we invoke to obtain monotones of the time evolution
%, which has also been employed in the context of resource theroies~\cite{marvian2014extending},
relies on the ``infinitesimal version'' of quantum data-processing inequalities. Distinguishability measures $D(\rho,\sigma) \ge 0$, defined over pairs of quantum states such that $D (\rho,\rho) = 0$, are said to obey the data-processing inequality if they are non-increasing under the joint action of a quantum operation on both arguments and play a central role in quantum information theory~\cite{witten2018mini,wilde2013quantum}. Since Markovian dynamics is fully characterized by its generator, we invoke an infinitesimal version of distinguishability measures, connecting with monotone Riemannian metrics in the space of quantum states. Such metrics have proven useful for the study of Markovian quantum dynamics, as for instance for the study of the mixing time and convergence rates~\cite{temme2010chi,kastoryano2012cutoff,kastoryano2013quantum}.

%, as well as various results relating to symmetric Markovian dynamics

%. In Ref.~\cite{lostaglio2017markovian}  constraints in the evolution of symmetric Markovian dynamics.

The paper is organized as follows. In \autoref{sec:monotones} we begin by first non-rigorously presenting the basic idea for establishing the correspondence between symmetries and monotones of the evolution. We then present the general construction, which is related to monotone Riemannian metrics and, after giving a simple example, we establish a connection with the Noether conserved quantities in open systems. We also briefly discuss how one can construct symmetries for the generator of the dynamics and the monotones. In \autoref{sec:dephasing}, we apply the method to dephasing generators and in \autoref{sec:Davies} for generators of the Davies form, extracting in both cases the qualitative features of the evolution with symmetry considerations. We conclude in \autoref{sec:summary}.

%, such as the time evolution $\mathcal E_t$ of open systems, 

\section{Monotones of the evolution from symmetries of the dynamics} \label{sec:monotones}

\subsection{Setting the stage}

We consider quantum systems described by a state $\rho \in \mathcal S(\mathcal H) \subseteq \mathcal B (\mathcal H) $, where the Hilbert space $\mathcal H \cong \mathbb C^{d}$ is finite dimensional\footnote{We use $\mathcal B (\mathcal H)$ to denote the space of (bounded) linear operators. $\mathcal S(\mathcal H)$ denotes the space of non-negative linear operators with unit trace.}. We assume that the dynamics is Markovian and time-homogeneous, i.e., 
\begin{align}\label{eq:diffential_equation}
\frac{d}{dt} \rho_t = \mathcal L (\rho_t)
\end{align}
where the generator of the dynamics $\mathcal L$, also known as the Lindbladian, is a time independent superoperator and hence can be expressed in the standard form
\begin{align} \label{eq:Lindblad_form}
\mathcal L(X) = -i \left[ H , X \right] + \sum_i \big( L_i X L_i^\dagger - \frac{1}{2} \big\{ L_i^\dagger L_i , X  \big\} \big)
\end{align}
(see, e.g., Ref.~\cite{breuer2002theory} for more details).
Equation~\eqref{eq:diffential_equation} gives rise to a one-parameter family of time-evolution superoperators
\begin{align}
\mathcal E_t = \exp \left( t \mathcal L \right)\,\;, \quad t\ge 0 \,\;,
\end{align}
forming a semigroup\footnote{I.e., $\mathcal E_0 = \mathcal I$ and $\mathcal E_t \mathcal E_{t'} = \mathcal E_{t+t'}$ for all $t,t' \ge 0$. This type of quantum dynamics is also called semigroup evolution.} in the space of Completely Positive and Trace Preserving (CPTP) maps, also known as quantum channels. We define as symmetries of a Lindbladian those superoperators that commute with $\mathcal L$, i.e.,
\begin{align}
\Symm(\mathcal L) \coloneqq \left\{ \mathcal M  \in \mathcal B \left( \mathcal B \left( \mathcal H \right) \right) \mid \left[ \mathcal M , \mathcal L \right] = 0  \right\} \,.
\end{align}

We first informally present the basic idea of how one can obtain monotones of the evolution from symmetries of the generator. Let us consider a distinguishability measure, i.e., a function $D(\rho,\sigma)$ over pairs of states with
\begin{subequations} \label{eq:D_properties_basic}
\begin{align}
D(\rho,\sigma) \ge 0  \quad \text{and} \quad D(\rho,\rho) = 0 \,,
\end{align}
such that it respects the data-processing inequality
\begin{align} 
D(\rho,\sigma) \ge D\left(\mathcal E (\rho), \mathcal E (\sigma) \right)
\end{align}
\end{subequations}
for all states $\rho$, $\sigma$ and quantum channels $\mathcal E$. Let us also define
\begin{align}\label{eq:D_with_s}
	f_{s}(\rho) = D(\rho, e^{s \mathcal M } \rho) \,\; , \quad \text{with }  \mathcal M \in \Symm \left(\mathcal L \right)\,\;,\end{align}
%Since $\mathcal M$ commutes with the time evolution, we will call any $\mathcal M \in \Symm(\mathcal L)$ a symmetry of the dynamics. In words, $f_s(\rho)$ 
which is a measure of the difference between a state $\rho$ and its variation\footnote{Notice that for $\exp(s \mathcal M) (\rho)$ to be a valid quantum state for $s\in [0,T) \subseteq \mathbb R_{\ge 0}$, the superoperator $\mathcal M$ should obey certain constraints, as for instance that $\mathcal M(\rho)$ is hermitian and traceless. We will later generalize this construction to include any $\mathcal M \in \Symm \left(\mathcal L \right)$.} generated by the symmetry $\mathcal M$. 

The data-processing inequality, together with the fact that $\mathcal M$ is a symmetry of the time evolution, implies that
\begin{align}\label{eq:fs_monotonicity}
	f_s(\rho) \ge f_s (e^{t \mathcal L } \rho)\,\; , \quad \forall t \ge 0 \,\;,
\end{align}
i.e., that $f_s$ is a monotone of the dynamics. However, to calculate $f_s$ for a finite value of the parameter $s$ one needs to first calculate the action of the exponential $\exp(s \mathcal M)$ on $\rho$, which in general is impractical. Instead, if the function $f_s$ is suitably well-behaved, one can examine its expansion
\begin{align} \label{eq:expansion}
f_s(\rho) = s^k h_{\mathcal M}(\rho) + O(s^{k+1}) \,, \quad k\in \mathbb N \,\;,
\end{align}
where the form of $h_{\mathcal M}$ depends on $D$ and $\mathcal M$.
Since the inequality \eqref{eq:fs_monotonicity} is valid for all $s \ge 0$, no matter how small, it follows that also $h_{\mathcal M}(\rho)$ is a monotone of the evolution
\begin{align} \label{eq:h_monotonicity}
h_{\mathcal M}(\rho) \ge h_{\mathcal M}( \mathcal E_t \left( \rho \right)) \,\;, \quad t\ge 0 \,\;.
\end{align}

Before proceeding to formalize the previous observation via basic tools from quantum information-geometry, let us consider an example. If the distinguishability measure is taken to be the relative $\alpha$-R\'enyi entropy with $\alpha = 1/2$, i.e.,
\begin{align}
D(\rho,\sigma) = S_{\frac{1}{2}}(\rho,\sigma) = - 2 \log\left( \Tr \left[ \sqrt \rho \sqrt \sigma   \right] \right)
\end{align}
then it is not hard to show that the expansion \eqref{eq:expansion} yields $k = 2$ and
\begin{align} \label{eq:Wigner-Yanase}
h_{\mathcal M} (\rho) = \sum_{ij}  \frac{\left| \braket{i|  \mathcal M (\rho) |j}\right|^2}{\left( \sqrt {p_i} + \sqrt{p_j} \right)^2} \,\;,
\end{align}
where we spectrally decomposed $\rho = \sum_{i} p_i \ket{i}\!\bra{i}$. A detailed derivation of Eq.~\eqref{eq:Wigner-Yanase} via elementray methods can be found in Appendix~\ref{sec:app:derivation}. The above quantity is well-known in quantum information-geometry as the Wigner-Yanase metric~\cite{hansen2008metric}, here evaluated on $\mathcal M \left(\rho \right)$.

%The concept of symmetry arises rather naturally in this construction. In Eq.~\eqref{eq:D_with_s}, the role of the superoperator $\mathcal L$ is to generate a variation of the state $\rho$. As long as this variation commutes with the Lindbladian $\mathcal L$, and hence with the time evolution $\mathcal E_t = \exp (t \mathcal L)$, one can instead choose
%\begin{align}
%f_{s}(\rho) = D(\rho, e^{s \mathcal M } \rho) \,\; , \quad \left[ \mathcal L, \mathcal M \right] = 0 \,, \quad  s \ge 0 \,\;,
%\end{align}
%implying that $g_{\mathcal M} (\rho)$ is a monotone of the evolution generated by $\mathcal L$.

\subsection{Monotones of the evolution and monotone Riemannian metrics}

%It might be clear at this point that an appropriate mathematical framework to systematically describe the resulting functions $g_\mathcal M$ is given by differential geometry.

In the previous section we discussed a way of obtaining monotones for a Markovian evolution given a distinguishability measure and a symmetry $\mathcal M$ of the generator, as in the example of Eq.~\eqref{eq:Wigner-Yanase}. This raises the question of how to systematically perform the expansion \eqref{eq:expansion} for a suitably wide class of distinguishability measures, a question that naturally leads to the consideration of monotone Riemannian metrics.

To see that, consider to the case where $\sqrt{D(\rho,\sigma)}$ is a distance function over the manifold $\mathcal S_{>0}(\mathcal H)$ of positive definite states
%, i.e., in addition to properties \eqref{eq:D_properties_basic} it is also symmetric, non-degenerate, and obeys the triangle inequality.
such that $\sqrt D$ arises from a Riemannian metric\footnote{I.e., the distance between two states is the length of the geodesic connecting them.} $g$. In that case, the expansion~\eqref{eq:expansion} yields $k=2$ and the function $h_{\mathcal M} (\rho)$ is nothing else that (squared) length of the tangent vector $\mathcal M(\rho)$ over the tangent space $T_{\rho} \,\mathcal S_{>0}(\mathcal H) $, i.e.,
\begin{align}
h_{\mathcal M} (\rho) = g_{\rho} \left(\mathcal M (\rho), \mathcal M (\rho) \right) \,\;.
\end{align}
More importantly, the monotonicity property~\eqref{eq:h_monotonicity} is satisfied if the metric is contractive under the action of quantum channels $\mathcal T$, namely
\begin{align} \label{eq:metrics_data_processing}
g_{\rho} (X , X) \ge g_{\mathcal T(\rho)} (\mathcal T(X) , \mathcal T (X) ) \,\;.
\end{align}

The aforementioned class of Riemannian metrics, called monotone metrics~\cite{morozova1991markov,petz1996monotone,petz1996geometries,petz1998contraction}, constitute the quantum analog of the Fisher metric for probability distributions (see also Ref.~\cite{bengtsson2017geometry} for an accessible introduction to the subject). While the latter is uniquely determined from the monotonicity property under classical stochastic maps (up to a normalization constant), the same does not hold for its quantum counterparts, which show a rich variety\footnote{In fact, monotone metrics are also closely related to generalized relative entropies~\cite{petz1986quasi}, as every monotone Riemannian metric arises from a generalized relative entropy~\cite{lesniewski1999monotone}. Hence the previous discussion can be generalized to $D$ corresponding to a relative entropy, as in our example~\eqref{eq:Wigner-Yanase}.}.

We include for completeness an elementary discussion of monotone Riemannian metrics in Appendix~\ref{sec:app:monotone_metrics}.
%, where we also recall how the above inequality connects with the monotonicity property.

For the purposes of this work, the crucial property of such quantities is that they satisfy the data-processing inequality, Eq.~\eqref{eq:metrics_data_processing}. This monotonicity property can be understood as a consequence of a key operator inequality, first proved in Ref.~\cite{lesniewski1999monotone}, which we repeat here due to its central importance for what follows.

\begin{theorem*}[Lesniewski and Ruskai~\cite{lesniewski1999monotone}]
\begin{multline} \label{eq:main_inequality}
	\Tr \left[ A^\dagger \left( \mr{\sigma} + \lambda \ml{\tau} \right)^{-1} (A) \right] \ge  \\ \Tr \left[  \mathcal E (A) ^\dagger  \left( \mr{\mathcal E (\sigma)} + \lambda \ml{\mathcal E(\tau)}\right)^{-1} \left[ \mathcal E(A) \right] \right] ,
\end{multline}
where $A \in \mathcal B(\mathcal H)$ is a linear operator, $\ml{\sigma} (X) \coloneqq \sigma X$ $(\mr{\tau}(X) \coloneqq X \tau)$ is the superoperator representing left (right) multiplication, $\mathcal E$ is a quantum channel and $\lambda \in \mathbb R_{\ge 0}$. The operators $\sigma,\tau \in \mathcal B_{>0} (\mathcal H)$ are positive definite, assuring that the superoperator inverses entering the inequality are well-defined, as well as that the resulting traces are non-negative.
\end{theorem*}

Let us now return to considering quantum Markovian dynamics generated by some time-independent Lindbladian $\mathcal L$. For our purposes, the quantum channel in the inequality~\eqref{eq:main_inequality} is specialized to the time evolution superoperator $\mathcal E_t \coloneqq \exp \left(t \mathcal L \right)$ for $t \ge 0$. Letting $\rho \in \mathcal S_{>0} (\mathcal H)$ be a full-rank state, we take
%in place of the general operator $A$ the expression
\begin{align}
A = \mathcal M (\rho) &\,, \quad  \text{where } \mathcal M \in \Symm\left(\mathcal L \right) \\
\sigma = \tau = \mathcal N (\rho) &\,, \quad  \text{where } \mathcal N \in \Symm\left(\mathcal L \right)
\end{align}
such that $\mathcal N (\rho) \in \mathcal B _{>0} (\mathcal H)$ is positive definite for all $\rho \in \mathcal S_{>0} (\mathcal H)$.
We will refer to any such superoperators $\mathcal M$ and $\mathcal N$ as symmetries of the dynamics.

The resulting inequality, due to the commutation relations $\left[ \mathcal M , \mathcal E_t  \right] = \left[ \mathcal N , \mathcal E_t  \right] = 0$, expresses the fact that the quantity
%$\mu_{\mathcal M}^{(\lambda)} : \mathcal S_+ (\mathcal H) \to \mathbb R_{\ge 0}$ with
\begin{align} \label{eq:monotones_general_M_N}
\cqf{\rho}{\mathcal M}{\mathcal N}{\lambda}  \coloneqq \Tr \!\left[  \mathcal M(\rho) ^\dagger \!\left( \mr{\mathcal N(\rho)} + \lambda \ml{\mathcal N(\rho)} \right)^{-1} \! \left[ \mathcal M(\rho) \right] \right]
\end{align}
%parameterized by $\lambda \ge 0$ and
satisfies
\begin{align} \label{eq:inequality_monotones_channel}
	\cqf{\rho}{\mathcal M}{\mathcal N}{\lambda} \ge \cqf{\mathcal E_t (\rho)}{\mathcal M}{\mathcal N}{\lambda}  \quad \text{for} \quad \lambda,t \ge 0.
\end{align}
We have shown the following.

\begin{proposition} \label{th:main_proposition}
If $\mathcal M, \mathcal N$ are symmetries of the dynamics and $\lambda \ge 0$, then the function
\begin{align*}
	\cq: \mathcal S_{>0} (\mathcal H) \to \mathbb R_{\ge 0} 
\end{align*}
defined in Eq.~\eqref{eq:monotones_general_M_N} is non-increasing under the time evolution generated by $\mathcal L$. 
\end{proposition}

In the rest of this paper, we will mainly focus on two cases.
\begin{enumerate}[(i)]
	\item $\mathcal N = \mathcal I$, in which case for simplicity we denote the resulting family of monotones as
	\begin{align} \label{eq:monotones_general}
	\cqt{\rho}{\mathcal M}{\lambda} = \Tr \left[  \mathcal M(\rho) ^\dagger \left( \mr{\mathcal \rho} + \lambda \ml{\mathcal \rho} \right) ^{-1} \left[ \mathcal M(\rho) \right] \right] \,\;.
	\end{align}
	\item If the dynamics admits a full-rank stationary state $\omega$, then $\mathcal N(X) = \Tr \left(X \right) \omega$ is a symmetry. In that case, it is convenient to denote directly the fixed state $\omega$ instead of the symmetry, i.e., write
	\begin{align} \label{eq:monotones_fixed_point}
	\cqf{\rho}{\mathcal M}{\omega}{\lambda} = \Tr \left[  \mathcal M(\rho) ^\dagger \left(\mr{\mathcal \omega} + \lambda \ml{\mathcal \omega}\right)^{-1}  \left[ \mathcal M(\rho) \right] \right] \,.
	\end{align}
\end{enumerate} 

The monotones $\cqt{\rho}{\mathcal M}{\lambda}$ can be also expressed in coordinates via spectrally decomposing the argument $\rho = \sum_i p_i \ket{i} \!\bra{i}$. Substituting, one gets
\begin{align} \label{eq:monotones_general_in_coordinates}
\cqt{\rho}{\mathcal M}{\lambda} = \sum_{ij} \frac{1}{\lambda p_i + p_j} \left| \braket{i|\mathcal M(\rho) | j}\right|^2 \,\;.
\end{align}
Similarly, for the case of $\cqf{\rho}{\mathcal M}{\omega}{\lambda}$ one can decompose $\omega = \sum_i q_i \ket{i_\omega} \!\bra{i_\omega} $ resulting in 
\begin{align}
\cqf{\rho}{\mathcal M}{\omega}{\lambda} = \sum_{ij} \frac{1}{\lambda q_i + q_j} \left| \braket{i_\omega|\mathcal M(\rho) | j_\omega}\right|^2 \,\;.
\end{align}

For Hamiltonian dynamics, $\mathcal L (X) = \mathcal K_H (X) \coloneqq -i \left[ H,X\right] $, the monotones \eqref{eq:monotones_general_M_N} are, in fact, conserved. This follows because unitary channels are invertible, hence for this case Eq.~\eqref{eq:inequality_monotones_channel} holds true for $t \in \mathbb R$. This forces monotones to maintain a constant value along the orbit.

For every family $\{\cq\}_\lambda$, one can consider convex combinations according to the measure $\mu(\lambda)$, namely
\begin{align} \label{eq:averaging_monotones}
\{\cq\}_\lambda \mapsto \int d\mu(\lambda) \cq  \,\;.
\end{align}
From this construction one obtains valid monotones, possibly admitting a convenient mathematical form for an appropriate choice of the measure (for explicit expressions of measures, see~\cite{hansen2008metric}). However, notice that the resulting functions do not impose any additional constraints compared to the ones from the parent functions\footnote{I.e., if the possibility of a transition $\rho \mapsto \sigma$ is ruled out by the inequality $\int d\mu(\lambda)\cqf{\rho}{\mathcal M}{\mathcal N}{\lambda} < \int d\mu(\lambda) \cqf{\sigma}{\mathcal M}{\mathcal N}{\lambda}$, then there exists a (non-zero measure) set of $\lambda$'s for which also $\cqf{\rho}{\mathcal M}{\mathcal N}{\lambda} <  \cqf{\sigma}{\mathcal M}{\mathcal N}{\lambda}$.}.

%$\cqf{\rho}{\mathcal M}{\mathcal N}{\lambda} \ge \cqf{\sigma}{\mathcal M}{\mathcal N}{\lambda}$  then also $\int d\mu(\lambda)\cqf{\rho}{\mathcal M}{\mathcal N}{\lambda} \ge \int d\mu(\lambda) \cqf{\sigma}{\mathcal M}{\mathcal N}{\lambda}$}

%$\llbracket \rho \rrbracket _{ \mathcal M, \mathcal N}^{(\lambda)}$}.

%$\cqf{\rho}{\mathcal M}{\mathcal N}{\lambda} \ge \cqf{\sigma}{\mathcal M}{\mathcal N}{\lambda}$  then also $\int d\mu(\lambda)\cqf{\rho}{\mathcal M}{\mathcal N}{\lambda} \ge \int d\mu(\lambda) \cqf{\sigma}{\mathcal M}{\mathcal N}{\lambda}$}.

%$\cqf{\rho}{\mathcal M}{\mathcal N}{\lambda} \ge \cqf{\sigma}{\mathcal M}{\mathcal N}{\lambda}$  then also $\int d\mu(\lambda)\cqf{\rho}{\mathcal M}{\mathcal N}{\lambda} \ge \int d\mu(\lambda) \cqf{\sigma}{\mathcal M}{\mathcal N}{\lambda}$

%Finally, we note that the class of monotones \eqref{eq:monotones_general_M_N} that we are considering here includes at least all constraints imposed by the case where $\sqrt{D}$ is a monotone distance arising from a Riemannian metric.
% and the case of $D$ being a generalized divergence \geo{check}.
%This is because, from the above theorem, one can recover as a special case the data processing inequality of the monotone metrics by use of standard mathematical results in integral representations of operator monotone functions~\cite{bhatia2013matrix}.
% that the monotonicity of the Riemanian metrics.

\subsection{A first example: Dephasing of a qubit} \label{sec:qubit_dephasing}

It might be useful at this stage to consider a simple example to illustrate the formalism. Let us analyze a two-level system with dephasing dynamics described by the Lindbladian
\begin{align}
\mathcal L (X) = -i g \left[ \sz , X \right] + \sz X \sz - X \,, \quad g \in \mathbb R\,\;.
\end{align}
The time evolution in terms of the Bloch vector representation of the state $\rho(t) = \frac{1}{2} \left( I + \bs v \cdot \bs \sigma \right)$ is, in cylindrical coordinates,
\begin{align} \label{eq:qubit_spiral_trajectory}
r_t = r_0 e^{-2t}\,,\quad  \phi_t = \phi_0 + 2 g t\,, \quad   z_t = z_0 \,\;.
\end{align}
That is, the Bloch vector lies onto a horizontal plane evolving inwards in a spiral motion.

Let us now illustrate how one can deduce the qualitative features of the evolution just by symmetry considerations, by use of Eq.~\eqref{eq:monotones_general}. Since the Hamiltonian $H = g \sz$ and the single Lindblad operator $L = \sz$ are both diagonal in the $\sz \coloneqq \ket{0}\!\bra{0} -\ket{1}\!\bra{1}$ eigenbasis, clearly the left multiplication superoperators $\ml{\ket{0}\!\bra{0}}$ and $\ml{\ket{1}\!\bra{1}}$ are symmetries of the Lindbladian. One immediately gets that the populations
\begin{align} \label{eq:qubit_dephasing_1}
\cqt{\rho}{\ml{\ket{i}\!\bra{i}}}{0} = \Tr\left( \rho \ket{i} \! \bra{i} \right) \,, \quad i = 0,1
\end{align}
are non-increasing. However, since the evolution is trace preserving, each of the populations is separately conserved. Notice that these are exactly the two linearly independent conserved quantities of the evolution predicted by Noether's theorem for Lindbladians, which we discuss momentarily. 

Now we consider again Eq.~\eqref{eq:monotones_general} but for the symmetry $\mathcal M (X) =  \mathcal K_{\sz}$ and $\lambda = 1$. Spectrally decomposing $\rho = p_+ P_+ + p_- P_-$ and invoking Eq.~\eqref{eq:monotones_general_in_coordinates}, we have
\begin{align}\label{eq:qubit_dephasing_2}
\cqt{\rho}{\mathcal K_{\sz}}{1} 
%&= \sum_{i,j \in \{ +,- \}} \left|  \left( \sz \right)_{ij} \right|^2 \frac{(p_i - p_j)^2}{p_i + p_j} \\
& = 2 \Tr \left( P_+ \sz P_- \sz  \right) (p_+ - p_-)^2 \nonumber  \\
&= 2 \left| \braket{0 | \rho | 1}\right|^2 \,\;,
\end{align}
that is, coherences are non-increasing under the time evolution.

The monotones from Eqs.~\eqref{eq:qubit_dephasing_1} and \eqref{eq:qubit_dephasing_2} jointly limit the set of states $R$ that are (possibly) reachable under time evolution. The aforementioned set consists of all Bloch vectors lying on horizontal disk with radius equal to $r_0$ (the radial distance of the initial Bloch vector from the $z$-axis.) 

In fact, the set $R$ of our example is as constrained as possible, in the sense that additional symmetry arguments (relying on the set $\Symm (\mathcal L)$) cannot restrict it more. To see this, first notice that the set of symmetries of our dephasing Lindbladian does not depend on the value of the parameter $g$, as long as $g \ne 0$. On the other hand, by varying $g$, all points in the bulk of $R$ can be reached by the trajectories of Eq.~\eqref{eq:qubit_spiral_trajectory} (for any fixed initial condition). This is because $g$ controls the frequency of the rotational motion, which can be made arbitrarily high. Therefore no more points in $R$ can be excluded by symmetry arguments.

%\begin{subequations}
%\begin{align}
%v_x &= e^{-2t} r \sin \theta \cos(\phi + 2 \omega t) \\
%v_y &= e^{-2t} r \sin \theta \sin(\phi + 2 \omega t) \\
%v_z &= r \cos \theta
%\end{align} 
%\end{subequations}

\subsection{Monotones imply Noether conserved quantities}

We will say that an operator $Y \in \mathcal B(\mathcal H)$ is conserved if
\begin{align}
\Tr\left[ Y^\dagger \exp(t \mathcal L) (\rho) \right]  = \text{const} \quad \forall \rho \text{ and } t\ge 0 \,\;.
\end{align}
We will also refer to such functions of the state as Noether conserved quantities, since they generalize the corresponding Hamiltonian construction. 
It is well-known that
\begin{align}
Y \text{ is conserved\quad} \Longleftrightarrow \quad  Y \in \Ker \mathcal L^*
\end{align}
(the adjoint is with respect to the Hilbert-Schmidt inner product).
The claim follows by noticing that the above trace can be expressed via the Hilbert-Schmidt inner product ($\braket{A,B} \coloneqq \Tr \left( A^\dagger B \right)$ for $A,B \in \mathcal B (\mathcal H)$) as 
\begin{align*}
\Tr\left[ Y^\dagger \exp(t \mathcal L) (\rho) \right] &= \braket { Y, \exp(t \mathcal L) (\rho) } \\ 
&= \braket {\exp(t \mathcal L^*) (Y),  \rho } .
\end{align*}

A consequence is that the number of linearly independent conserved quantities equals the dimension of the subspace spanned by the steady states of the evolution, a fact also noted in \cite{albert2014symmetries}. This observation follows from the identity $\dim \Ker(\mathcal L) = \dim \Ker(\mathcal L^*) $. In particular, if there is a unique fixed point of the evolution then all conserved quantities trivialize, in the sense that they are necessarily proportional to the trace of the (time-evolved) state, since always $I \in \Ker{\mathcal L^*}$.
%This is because a unique fixed point implies $\dim \Ker(\mathcal L^*) = 1$. However, since always $I \in \Ker{\mathcal L^*}$, all conserved quantities are proportional to the trace of the state, hence do not  carry any useful information.

On the other hand, the monotones $\cqt{\rho}{\mathcal M}{\lambda}$ impose non-trivial, in general, constraints on the reachable states of the evolution, for instance by taking $\mathcal M = \mathcal L$ (a specific example is given later in \autoref{fig:Davies}). What is more, if a steady state $\omega$ is known (regardless whether it is unique or not), then also monotones of the form $\cqf{\rho}{\mathcal M}{\omega}{\lambda}$ can be utilized, imposing additional constraints.

A natural question to be asked is whether the constraints imposed by the Noether conserved quantities are included in the monotones arising from \autoref{th:main_proposition} or not. The answer is affirmative.

\begin{proposition}
If the Lindbladian admits a full-rank stationary state, then for each conserved operator $Y$ there is an appropriate choice of the symmetries $\mathcal M, \mathcal N$ in the family of monotones $\cqf{\rho}{\mathcal M}{\mathcal N}{0}$ that implies the conservation.

\end{proposition}
Notice, however, the converse is not true; constraints imposed by the monotones $\cqf{\rho}{\mathcal M}{\mathcal N}{0}$ cannot necessarily be inferred from conservation laws.

\begin{proof}
Let us consider some conserved $Y>0$. The superoperator $\mathcal M (\rho) = \Tr \left( Y \rho \right) \omega$, for $\omega$ a full-rank stationary state of the evolution, is a symmetry of the dynamics. Therefore $\cqf{\rho}{\mathcal M}{\omega}{0} = \Tr \left( Y \rho \right)^2$ is non-increasing under the time evolution, hence also $\Tr \left( Y \rho \right)$ is non-increasing.

On the other hand, since $\omega$ is a full-rank steady state and $\Tr \left( Y \rho \right) > 0$, also $\cqf{\rho}{\omega}{\mathcal M}{0} = \left[ \Tr \left( Y \rho \right) \right]^{-1}$ is a valid non-increasing monotone. As a result, $\Tr \left( Y \rho \right)$ has to be a constant of the evolution.

To complete the proof, we need to show that the constraints imposed by conserved $Y \in \mathcal B_{>0}(\mathcal H)$ are the same as the ones imposed by arbitrary $Y \in \mathcal B(\mathcal H)$. Indeed, consider some (possibly non-hermitian) $Y \in \Ker(\mathcal L^*)$. Since $\mathcal L^*$ preserves hermiticity, then both the hermitian and anti-hermitian parts of $Y$ are separately conserved i.e., conservation of $Y$ amounts to conservation of the hermitian operators $(Y + Y^\dagger)/2$ and $(Y - Y^\dagger)/(2i)$. Finally notice that, although these two operators might fail to be positive, due to unitality of $\mathcal E^*_t$ it holds that $I \in \mathcal \Ker \left( \mathcal L^* \right)$. Hence, if a hermitian $Y$ is conserved, then also $Y + a I $ is conserved, which can always be made positive for large enough $a$. Notice that this amounts to just adding the constant $a$ to the value of the Noether conserved quantity.

%This is true because $\mathcal L^*$ preserves hermiticity and therefore $Y \in \Ker(\mathcal L^*)$ also implies $Y^\dagger \in \Ker(\mathcal L^*)$, allowing to trade a possibly non-hermitian $Y$ for its . 

The above demonstrate that the monotone family $\cqf{\rho}{\mathcal M}{\mathcal N}{0}$ imposes at least as many constraints as conserved quantities.
\end{proof}

\subsection{Symmetries of the generator and the monotones}

We now very briefly address the question of how one can find symmetries of a Lindbladian, beyond those guessed by inspection.
%, and recall some useful facts.
Determining explicitly the entire set of symmetries $\Symm \left(\mathcal L \right)$ of a Lindbladian is usually too complicated to be of practical importance. Instead, when a representation of the Lindbladian in Lindblad operators is known (i.e., a decomposition as in Eq.~\eqref{eq:Lindblad_form}), it might be convenient to look at operators simultaneously commuting with all elements of the set $S = \{ H \}  \cup \big\{ L_i,L_i^\dagger \big\}_i$. If $\mathcal A$ is the algebra generated by $S$, then the element of the algebra generated by $\{ \ml{X} , \mr{X}\}_{X \in \mathcal A'}$ belong to $\Symm \left( \mathcal L \right)$ (here $\mathcal A'$ denotes the commutant of the algebra\footnote{I.e., the operator subalgebra of $\mathcal B(\mathcal H)$ consisting of all (and only) the elements that commute with every element of $\mathcal A$.}). We will use this fact to analyze dephasing generators later in the paper.

Steady states $\omega$ give rise to symmetries, for instance via maps $\mathcal N (X)=  \Tr (X) \omega$, as in Eq.~\eqref{eq:monotones_fixed_point}. Although $\Ker \left( \mathcal L \right)$ is not in general closed with respect to operator multiplication, it was shown in Ref.~\cite{idel2013structure}  that it forms an Euclidean Jordan Algebra. More specifically, if the limit $F \coloneqq \lim_{t \to \infty} \exp\left( t \mathcal L \right) I$ exists and is full-rank, then for any $A,B$ fixed points of the evolution it holds that
\begin{align}
A \bullet_{\!{}_F}\!\! B \coloneqq \frac{1}{2} \left( A_F B_F + B_F A_F \right) 
\end{align}
is also a fixed point\footnote{Notice that any operator $X \in \Ker\left( \mathcal L \right)$ can be decomposed into (up to four) steady states by considering the positive/negative part of $\frac{1}{2}(X + X^\dagger)$ and $\frac{1}{2 i}(X - X^\dagger)$. Each of the latter, after proper normalization, forms a steady state. This follows from the fact that the corresponding quantum channel $\mathcal E_t$ preserves positivity.}, where $X_F \coloneqq F^{-1/2} X F^{-1/2}$.

In addition, in Ref.~\cite{novotny2018quantum} the authors constructed a family of maps that are bijections over the subspace of fixed points of the dynamics. More specifically, the mappings can be constructed with the input of an operator monotone function~\cite{bhatia2013matrix} and two full-rank steady states. The above facts can be of use to generate more fixed points, and hence symmetries, out of a set of known ones.

Finally, we address the question of how to construct monotones that themselves possess some symmetry. Let us consider a unitary representation $U_g$ of a group $G$ such that $\mathcal U_g = U_g (\cdot) U_g^\dagger$ is a symmetry of the Lindbladian for all $g \in G$. Then one can define a (left) group action on the space of monotones via
\begin{align}
\cqf{\cdot}{\mathcal M}{\mathcal N}{\lambda} \xmapsto{\,g\,} \cqf{\cdot}{\mathcal U_g \mathcal M}{\mathcal N}{\lambda} = \cqf{\cdot}{ \mathcal M}{\mathcal U_{g^{-1}} \mathcal N}{\lambda} \,\;.
\end{align}
From the latter, it is straightforward to construct monotones that are invariant under the action by invoking the (Haar) group average, i.e., consider
\begin{align} \label{eq:covariant_monotone}
Q_{\mathcal M,\mathcal N}^{(\lambda)}(\rho) \coloneqq \int d \mu_{\text{Haar}}(g) \cqf{\rho }{\mathcal U_g \mathcal M}{\mathcal N}{\lambda}  \,\;.
\end{align}
%= \int d \mu(U_g) \, \mathcal U_g (\rho^2) \omega^{-1}
For instance,
\begin{align}
Q_{\mathcal M,\omega}^{(0)}(\rho) = \Tr \left[ \mathcal P \left( \mathcal M(\rho)^\dagger \mathcal M (\rho)\right) \omega^{-1} \right] \,\;,
\end{align}
where $\mathcal P (X) \coloneqq  \int d \mu_{\text{Haar}}(g) \mathcal U_g (X) $ is the projector superoperator onto the $G$-invariant subspace, i.e., for any operator $X$ it holds $U_g  \mathcal P(X) = \mathcal P(X) U_g$ $\forall g$. The projector $\mathcal P$ can be constructed explicitly in terms of the irreducible representations of $\{ U_g \}_g$ using standard techniques from theory of noiseless subsystems~\cite{knill2000theory,zanardi2000stabilizing}.

%In fact, one can consider the (usually simpler to construct) algebra $\mathcal A$ generated by the set $S$ and $U_g \in \Comm \left( \mathcal A \right)$ \geo{how to say $U_g$ is ``maximal''?}. In that case, Eq.~\eqref{eq:covariant_monotone} becomes
%\begin{align}
%Q(\rho) = \Tr \left( \mathcal P_{\mathcal A} \left(\rho^2 \right) \omega^{-1} \right) \,\;,
%\end{align}
%where $\mathcal P_{\mathcal A}$ is the projector over $\mathcal A$. This allows the use of standard techniques from representation theory \geo{What to cite?}, where $\mathcal P_{\mathcal A}$ admits explicit expressions in terms of the irreducible representations of $\mathcal A$.

\section{Dephasing generators} \label{sec:dephasing}

Here we consider Lindblad generators of the general form~\eqref{eq:Lindblad_form} such that all elements of the set $S = \{ H \}  \cup \big\{ L_i,L_i^\dagger \big\}_i$ mutually commute with each other. We will refer to this class as dephasing generators. Notice that this family is a generalization of the qubit example of  \autoref{sec:qubit_dephasing}.

%Dephasing Lindbladians are (Hilbert-Schmidt) normal operators, hence they admit a complete family of eigenvectors.
Dephasing Lindbladians are both unital and Hilbert-Schmidt normal, as can be inferred with a direct calculation invoking the standard form~\eqref{eq:Lindblad_form} of the generator. Hence by the spectral theorem they admit a complete family of eigenoperators. The corresponding time evolution can be formally expressed via resorting to the (maximal) projectors onto the joint eigenspaces of all elements in $S$. Let us denote as $\{ P_i \}_{i=1}^n$ ($n \le d$) this complete family of orthogonal projectors, and also as $\mathcal P_{ij} (X) \coloneqq P_i X P_j$ the corresponding family of (Hilbert-Schmidt orthogonal) projector superoperators. Then, for dephasing generators, the time evolution can be written as
\begin{align}
\exp\left( t \mathcal L \right)  = \sum_{ij} e^{\lambda_{ij} t} \mathcal P_{ij}  \,\;.
\end{align}
Invoking once again the standard form of the generator~\eqref{eq:Lindblad_form}, it follows that the eigenvalues satisfy  $\lambda_{ij} = \lambda_{ji}^*$ and $\lambda_{ii} = 0$ $\forall i$, while $\Real \left( \lambda_{ij} \right) < 0 $ for $i \ne j$.

The last inequality expresses that fact that there are no steady states with support over any of the eigenspaces $\mathcal P_{ij}$ with $i \ne j$. This is true since, for unital Lindbladians, $\Ker \left( \mathcal L \right)$  coincides with $\mathcal A'$~\cite{kribs2003quantum}, where $\mathcal A$ is the (abelian) algebra generated by the set $S$. In our case, this commutant is exactly given by operators in the range of $\sum_i \mathcal P_{ii}$.

Qualitatively, the evolution dictates that the diagonal elements $\sum_i \mathcal P_{ii}(\rho)$ are conserved while the off-diagonal parts $\sum_i \mathcal P_{ij} (\rho)$ ($i \ne j$) decay, possibly with oscillations, which is the defining characteristic of dephasing.

Let us now extract the same qualitative information from a symmetry analysis. Clearly, $\left[ \ml{P_i} , \mathcal L \right] = 0$ $\forall i$ hence
\begin{align}
\cqt{\rho}{\ml{P_i}}{0} = \Tr\left( \rho P_i \right) \,, \quad \forall i 
\end{align}
are non-increasing. On the other hand, the evolution is trace-preserving, therefore all $P_i$ are, in fact, conserved. These are all the linearly independent Noether conserved quantities.

In addition, $\ml{P_i}\mr{P_j}$ is a symmetry of the evolution since all elements in $S$ mutually commute. Combining this with the fact that the evolution is unital, hence $\omega = I$ is a fixed point, we directly get from Eq.~\eqref{eq:monotones_fixed_point} that
\begin{align}
\cqf{\rho}{\ml{P_i}\mr{P_j}}{I}{0} = \Tr \left( P_i \rho P_j \rho \right)
\end{align} 
is non-increasing. This corresponds to the decay of the $(i,j)$-coherences of the state $\rho$ and is a generalization of Eq.~\eqref{eq:qubit_dephasing_2}.

We have shown the following general fact.

\begin{proposition}
Let $\mathcal L$ be a Lindbladian.
\begin{enumerate}[(i)]
\item If $\ml{P}$ is a symmetry of $\mathcal L$, then $\Tr \left( \rho_t P^\dagger P \right)$ is a non-increasing function of time.
\item If  $\mathcal L$ is unital and $\ml{P}\mr{Q}$ is a symmetry, then $\Tr \left( P^\dagger P \rho_t Q^\dagger Q  \rho_t \right)$ is a non-increasing function of time.
\end{enumerate}
For $P,Q$ orthogonal projectors, then $(i)$ corresponds to decaying population and $(ii)$ to decaying coherence.
\end{proposition}

\section{Davies generators}  \label{sec:Davies}

\subsection{Preliminaries}

\begin{figure}[t]
\centering
\includegraphics[width=0.85\columnwidth]{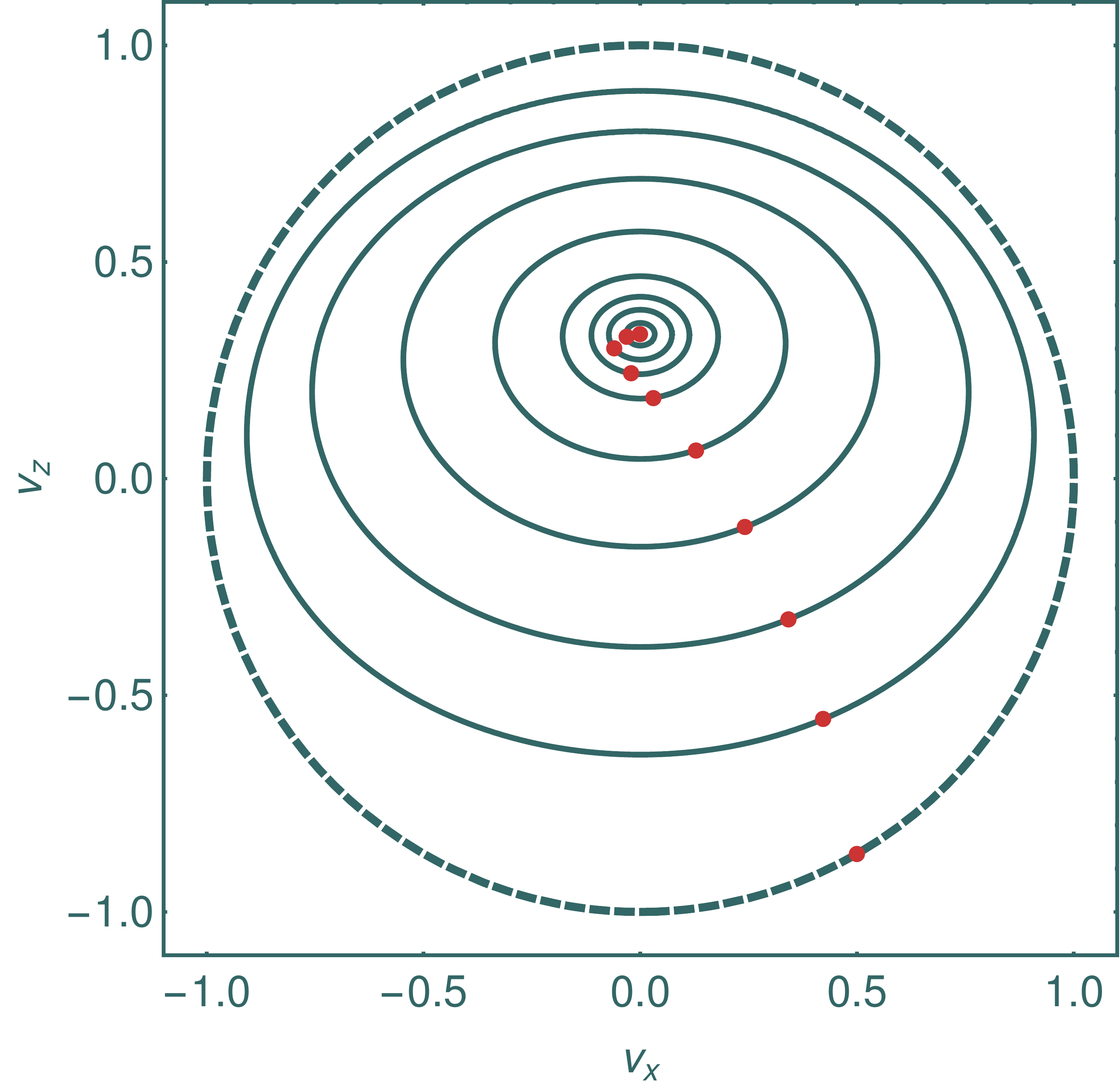}
\caption{\textbf{Constraints imposed on the time evolution by considering as symmetry the Lindbladian $\mathcal L$ itself}. In this example $\mathcal L$ is a single qubit Davies generator with $H = \sz$ and jump operators $L_1 = \sqrt{2} \sigma^+$, $L_2 = \sigma^-$. We depict in Bloch coordinates contours separating the allowed (inside) from the forbidden (outside) regions as predicted by the monotone $\cqt{\rho}{\mathcal L}{1/2}$ for various $\rho$. The monotone is symmetric under rotations around the $z$-axis, so we only plot an $x-z$ slice of the Bloch ball. The corresponding states $\rho$ are marked with a point and are chosen to correspond to instances of the actual trajectory (as projected onto the $x-z$ plane) of the initially pure state $ \bs v = (1/2, 0 ,-\sqrt{3}/2)$.} \label{fig:Davies}
\end{figure}

A physically significant class of Lindbladians is provided by the Davies generators~\cite{davies1974markovian} (for a more modern treatment, see also Refs.~\cite{alicki2007quantum,roga2010davies,rivas2012open}). This family arises from a first-principles derivation under the main assumption that the interaction between the system and its environment is weak and gives rise to steady states that are thermal. Davies generators have the following properties:
\begin{enumerate}[(i)]
\item The Hamiltonian part commutes with the dissipative part, i.e., $\mathcal L = \mathcal K_H + \mathcal D$ with 
\begin{subequations} \label{eq:Davies_general_all}
\begin{align}
\left[ \mathcal K_H  , \mathcal D  \right] = 0 \,\;.
\end{align}
\item It satisfies the quantum detailed balance condition, i.e., there exists a state $\tau$ such that

\begin{align} 
\mathcal D \mr{\tau} &= \mr{\tau} \mathcal D^* \,\;,  \label{eq:detailed_balance_dissipative} \\
\mathcal K_H \left(\tau \right) &= 0 \,\;. 
\end{align}
\end{subequations}
\end{enumerate} 
For the scope of this paper, we will refer to any Lindbladian that satisfies the conditions of Eqs.~\eqref{eq:Davies_general_all} as a Davies generator.

Let us recall a few useful mathematical facts regarding the detailed balance condition~\cite{alicki2007quantum,venuti2016dynamical}. First of all, since $I \in \Ker \left( \mathcal D^* \right)$, then Eq.~\eqref{eq:detailed_balance_dissipative} implies that $ \tau$ is a fixed point of the evolution. The detailed balance condition can be equivalently understood as the requirement that $\mathcal D^*$ is hermitian with respect to the (possibly degenerate) scalar product
\begin{align} \label{eq:Davies_scalar_product}
\braket{A,B}_{\tau} \coloneqq \Tr \left( \tau A^\dagger B \right) \,\;.
\end{align}
Assuming that $\tau$ is full rank, the detailed balance condition Eq.~\eqref{eq:detailed_balance_dissipative} can be recast in a yet another form, namely $\mathcal D$ being hermitian with respect to the scalar product $\braket{\cdot, \cdot}_{\tau^{-1}}$. The last claim follows by multiplying both from the left and the right Eq.~\eqref{eq:detailed_balance_dissipative} with $\mr{\tau}^{-1}$. 

Let us now express the general form of the time evolution of a Davies generators. Notice that Davies generators are always normal operators, hence admit a complete family of eigenoperators. This is because, by assumption, $\mathcal D$ is hermitian with respect to the scalar product \eqref{eq:Davies_scalar_product} (for $\tau^{-1}$) and it commutes with the Hamiltonian part (which is also anti-hermitian with respect to the same scalar product). Let us formally denote as $\{ \mathcal P_i \}_i$ the complete family of hermitian (with respect to $\braket{\cdot , \cdot}_{\tau^{-1}}$) superoperator projectors. Then the general solution has the form
\begin{align} \label{eq:Davies_time_evolution}
\exp \left(  t \mathcal L \right) = \sum_i  e^{\lambda_i t}  \mathcal P_i
\end{align}
with $\mathcal P_i \mathcal P_j = \delta_{ij} \mathcal P_i$ and $\sum_i \mathcal P_i = \mathcal I$.

\subsection{Davies generator: Single qubit}

Let us consider a qubit system whose Lindbladian is described by $H = \sz$ together with two Lindblad operators $\{ a \sigma^+ , b\sigma^-\}$. The unique steady state $\tau = \frac{1}{2} \left( I + \bs w \cdot  \bs \sigma \right)$ has a corresponding Bloch vector lying along the $z$-axis with $w_z =  \frac{a^2 - b^2}{a^2 + b^2} $. One can easily check directly that Eqs.~\eqref{eq:Davies_general_all} are satisfied, hence $\mathcal L$ is indeed a valid Davies generator. Setting $g = |a|^2 + |b|^2$, the dynamical evolution of the Bloch vector is, in cylindrical coordinates,
\begin{multline} \label{eq:qubit_Davies_trajectory}
r_t = r_0 e^{-g t / 2}\,,\;\,  \phi_t = \phi_0 + 2 t\,, \;\,  \\ z_t = (1 - e^{- g t}) w_z + e^{-g t} z_0 \,\;.
\end{multline}

Let us analyze some of its symmetries now. By definition of the Davies generator, $\mathcal K_{H} = \mathcal K_{\sz}$ is a symmetry, therefore Eq.~\eqref{eq:qubit_dephasing_2} is a valid monotone, indicating that coherences can only decay. This corresponds to the inward spiral motion of the trajectory.

Regarding the $z$-evolution of the Bloch vector towards the fixed point, let us assume that $\tau$ is known to be a fixed state of the evolution (but we do not explicitly assume that it is the unique fixed state). Then by inspection one can verify that $\mathcal M (X) = \braket{\sz , X}_{\tau^{-1}} \sigma_z$ is a symmetry of the evolution. Therefore
\begin{align} \label{eq:qubi_Davies_z_monotone}
\cqf{\rho}{\mathcal M}{\tau}{0} = \left| \Tr \left( \frac{1}{\tau} \sz \rho \right) \right|^2 \Tr \left( \tau^{-1} \right)
\end{align} 
is non-increasing, implying that also $\left| \Tr \left( \tau^{-1} \sz \rho \right) \right| $ is a monotone. In the Bloch representation, the last quantity equivalently expresses the fact that $\left| v_z - w_z \right|$ is non-increasing, i.e., the $z$-component of the Bloch vector monotonically approaches the steady state.

Finally, we numerically investigate in \autoref{fig:Davies} a example for the constraints imposed on the qubit time evolution just by considering as symmetry the Lindbladian itself.

\subsection{Davies generators: General remarks}

Let us make a few observations regarding monotones and Davies generators. First of all, notice that monotones \eqref{eq:monotones_fixed_point} for $\omega = \tau$ (the detailed balance fixed point) and $\lambda = 0$ are directly related to the Davies inner product
\begin{align}
\cqf{\rho}{\mathcal M}{\tau}{0} = \braket{\mathcal M (\rho), \mathcal M (\rho)}_{\tau^{-1}} \,\;.
\end{align}
In particular, the case $\mathcal M = \mathcal I$ expresses the fact that
\begin{align}
\cqf{\rho}{\mathcal I}{\tau}{0} = \Tr\left( \frac{1}{\tau} \rho^2 \right)
\end{align}
is non-increasing under the time evolution, which is a generalization of the well-known fact that the purity $\Pur(\rho) \coloneqq \Tr (\rho^2)$ is non-increasing under unital dynamics. Equivalently, the distance
\begin{align}
\left\| \rho_t - \tau  \right\|^2_{\tau^{-1}}  =  \braket{\rho_t - \tau, \rho_t - \tau}_{\tau^{-1}}
\end{align}
is non-increasing.

One can easily generalize the construction of the monotone \eqref{eq:qubi_Davies_z_monotone}. Each of the projectors in Eq.~\eqref{eq:Davies_time_evolution} can be written as $\mathcal P_i (X) = \sum_ {k=1}^{d_i} \braket{Y_{i,k} , X}_{\tau^{-1}} Y_{i,k}$, where $\{ Y_{i,k} \}_{i,k}$ is a complete orthonormal family of eigenoperators. Hence, for any such operator, $\mathcal M (X) = \braket{Y , X}_{\tau^{-1}} Y$ is clearly a symmetry of the dynamics, and because for these symmetries
\begin{align}
\cqf{\rho}{\mathcal M}{\tau}{0} = \left| \Tr \left( \frac{1}{\tau} Y^\dagger \rho \right) \right|^2
\end{align}
the functions $\left| \Tr \left( \tau^{-1} Y^\dagger \rho \right) \right|$ are monotones, generalizing Eq.~\eqref{eq:qubi_Davies_z_monotone}.

\section{Discussion and outlook} \label{sec:summary}

Identifying symmetries and their consequences in physical systems is an important task for various fields of physics. The present manuscript constitutes an attempt to provide a simple correspondence between symmetries in the generators of quantum Markovian dynamics and monotones of the corresponding evolution. More specifically, we presented a construction that assigns to every pair of symmetries of the generator a one-parameter family of functions over quantum states that are non-increasing under the time evolution. Such monotonic functions can be employed in order to identify states that are non-reachable by the evolution. The construction utilizes powerful tools from quantum information-geometry, mainly from the theory of monotone Riemannian metrics. Finally, we have demonstrated how one can deduce from symmetries the qualitative features of the evolution for the prototypical cases of dephasing and Davies generators. 

%The latter can be considered as the differential-geometric version of data processing inequalities.

%Symmetries in open quantum systems have been widely considered in the context of quantum resource theories~\cite{chitambar2019quantum}, specifically in the theory of~\cite{marvian2013theory}. There, given a group $G$, one investigates the possibility of state transitions under the class of quantum operations that are symmetric with respect to a unitary representation of $G$, i.e., the operations such that $\left[ \mathcal E , \mathcal U_g \right] = 0$ $\forall g \in G$.

%In Ref.~\cite{marvian2014extending}, the authors established the fact that for closed system dynamics the Noether conserved quantities are inadequate to capture all constraints regarding mixed state transitions under $G$-symmetric dynamics. On the other hand, asymmetry monotones can provide conserved quantities that are independent of the Noether ones. \geo{cite also Lostaglio}
%In the context of asymmetry, monotones arising from monotone metrics have also been applied in~\cite{takagi2019skew}.

One can easily rephrase the question addressed in the present paper to fit in the framework of quantum resource theories. Given some fixed Lindbladian, one can define as free operations those associated with the time evolution $\{ \mathcal E_t \coloneqq \exp\left( t \mathcal L \right) \}_{t\ge 0 }$. Our main result, \autoref{th:main_proposition}, provides a family of monotones for each pair of symmetries of the generator. However, even if all possible symmetries are considered, the resulting families of monotones are not in general complete, i.e., they do not exclude all states that are non-reachable by the evolution.
%\footnote{Notice that any such complete family cannot be solely a function of $\Symm \left(\mathcal L \right)$, as becomes apparent from the example of \autoref{sec:qubit_dephasing}.}.
It remains an open question to find a general construction for a complete family of monotones.
%Notice, however, that such a resource theory does not fall within the framework of asymmetry \geo{true?}.

We hope that the present work demonstrates yet another way in which ideas and tools from quantum information theory can provide useful insights to the analysis of quantum dynamics.

\begin{acknowledgments}
G.S.~is thankful to \'{A}.M.~Alhambra, N.~Anand, L.~Campos Venuti, A.~Hamma,  S.~Montangero, and N.A. Rodr\'{i}guez Briones for the interesting discussions and to Th.~Apostolatos and P.J.~Ioannou for their teachings. P.Z.~acknowledges partial support from the NSF award PHY-1819189.
\end{acknowledgments}

%\printbibliography

%\bibliographystyle{apsrev4-2}
%\bibliographystyle{unsrtnat}
\bibliographystyle{unsrturl}
\bibliography{refs}

\onecolumn
\newpage

\appendix

\section{Derivation of Eq.~\eqref{eq:Wigner-Yanase}} \label{sec:app:derivation}

We include below an explicit derivation of Eq.~\eqref{eq:Wigner-Yanase}, i.e., we calculate the expansion of the  $\frac{1}{2}$-R\'enyi entropy $S_{\frac{1}{2}} (\rho_0,\rho_s) = - 2 \log\left( \Tr \left[ \sqrt {\rho_0} \sqrt {\rho_s   } \right] \right)$ for $\rho_s = \exp\left(s \mathcal M \right) (\rho_0)$ to the first non-vanishing order in~$s$, assuming that the state $\rho_0$ is full-rank.

Setting $\chi_s = \sqrt{\rho_s}$, we have
\begin{align} \label{eq:app_trace_exp}
\Tr \left( \sqrt {\rho_0} \sqrt {\rho_s   }\right) = \braket{ \chi_0 , \chi_s } = 1 + s \braket{\chi_0 , \dot{\chi}_0}
 + \frac{s^2}{2} \braket{\chi_0 , \ddot{\chi}_0} + O\big(  s^3 \big) \,\;.
\end{align}
Moreover, from $\rho_s = \chi_s^2$ one finds
\begin{align*}
\dot{\rho_s} = \chi_s \dot{\chi}_s + \dot{\chi}_s \chi_s = \left( \ml{\chi_s} + \mr{\chi_s} \right) \dot{\chi}_s \coloneqq \mathcal A_{\chi_s} \left( \dot{\chi}_s \right) \,\;,
\end{align*}
hence also
\begin{gather*}
\dot{\chi}_s = \mathcal A_{\chi_s}^{-1} \left( \dot{\rho}_s \right) \,\;, \\ 
\ddot {\chi}_s = \mathcal A_{\chi_s}^{-1} \left( \ddot {\rho}_s \right) + \frac{d {\mathcal A}_{\chi_s}^{-1}}{ds} \left( \dot{\rho}_s \right)
 \,\;.
\end{gather*}
Using the above expressions we can calculate the terms in the expansion~\eqref{eq:app_trace_exp}. We have
\begin{align*}
\braket{\chi_s , \dot{\chi}_s} = \frac{1}{2}\braket{ {\mathcal A}_{\chi_s} (I),   \mathcal A_{\chi_s}^{-1} \left( \dot{\rho}_s \right) } = \frac{1}{2} \Tr \left(  \dot{\rho}_s \right) = 0
\end{align*}
hence the term linear in $s$ does not contribute. The quadratic term has two parts,
\begin{align*}
\braket{\chi_s , \ddot{\chi}_s}  = \braket{\chi_s ,\mathcal A_{\chi_s}^{-1} \left( \ddot {\rho}_s \right) }  + \braket{\chi_s , \frac{d {\mathcal A}_{\chi_s}^{-1}}{ds} \left( \dot{\rho}_s \right)}
\end{align*}
but one of them vanishes
\begin{align*}
\braket{\chi_s ,\mathcal A_{\chi_s}^{-1} \left( \ddot {\rho}_s \right) } = \frac{1}{2} \Tr \left( \ddot {\rho}_s   \right) = 0 \,\;.
\end{align*}
To calculate the remaining term notice that
\begin{align*}
{\mathcal A}_{\chi_s} =  {\mathcal A}_{\chi_0} + s \left( \ml{{ \dot\chi}_0} + \mr{{\dot \chi}_0} \right) + O\big( s^2 \big) = \left( \mathcal  I + s {\mathcal A}_{{\dot\chi}_0} {\mathcal A}_{\chi_0} ^{-1} \right) {\mathcal A}_{\chi_0}  + O \big( s^2 \big)
\end{align*}
and therefore
\begin{align*}
{\mathcal A}_{\chi_s} ^{-1} =  {\mathcal A}_{\chi_0} ^{-1} \left( \mathcal  I + s {\mathcal A}_{{\dot\chi}_0} {\mathcal A}_{\chi_0} ^{-1}\right)^{-1} + O \big( s^2 \big) \,\;,
\end{align*}
which gives
\begin{align*}
\frac{d {\mathcal A}_{\chi_s}^{-1}}{ds} \Bigg|_{s = 0}  =  -  {\mathcal A}_{\chi_0} ^{-1} {\mathcal A}_{{\dot\chi}_0} {\mathcal A}_{\chi_0} ^{-1} \,\;.
\end{align*}
Finally, this equation implies
\begin{align*}
\braket{\chi_0 , \ddot{\chi}_0} = - \frac{1}{2} \Tr \left[   {\mathcal A}_{{\dot\chi}_0} \left(  \dot{\chi}_0   \right) \right] = - \Tr \left[ \left(   \dot{\chi}_0  \right)^2 \right]
\end{align*}
which completes the expansions
\begin{align*}
\Tr \left( \sqrt {\rho_0} \sqrt {\rho_s   }\right) = 1 - \frac{s^2}{2}  \Tr \left[ \left(   \dot{\chi}_0  \right)^2 \right]
\end{align*}
and hence
\begin{align*}
S_{\frac{1}{2}} (\rho_0,\rho_s)  = s^2 \Tr \left[ \left(   \dot{\chi}_0  \right)^2 \right]   + O \big( s^3 \big) \,\;.
\end{align*}
The form~\eqref{eq:Wigner-Yanase} follows by using once again the fact that $\dot{\chi}_0 = \mathcal A_{\chi_0}^{-1} \left( \dot{\rho}_0 \right) $ and via expressing $\rho_0 = \sum_i p_i \ket{i}\!\bra{i}$.

\section{Monotone Riemannian metrics and Eq.~\eqref{eq:main_inequality}} \label{sec:app:monotone_metrics}

%In order to keep the present paper as self-contained as possible, 

Here we recall for completeness some well-known facts about monotone Riemannian metrics and discuss how the key inequality~\eqref{eq:main_inequality}, obtained by Lesniewski and Ruskai in~\cite{lesniewski1999monotone}, is related to the monotonicity property Eq.~\eqref{eq:metrics_data_processing}.

% its connection with generalized relative entropies.  \geo{cite}

Monotone Riemannian metrics over $\mathcal S_{>0}(\mathcal H)$, whose defining property is the data-processing inequality Eq.~\eqref{eq:metrics_data_processing}, were first classified completely by Petz~\cite{petz1996monotone,petz1996geometries}, who extended an earlier result by Morozova and Chentsov~\cite{morozova1991markov}. In short, every such metric admits the form
\begin{subequations}
\begin{align} \label{eq:metric_k}
g_{\rho} (X, Y)  =  \braket{X , \mathcal K_{\rho}^{-1} (Y)}
\end{align}
for $X,Y \in T_\rho \, \mathcal S_{>0}(\mathcal H)$ (i.e., are hermitian and traceless operators) and
\begin{align} 
\mathcal K_{\rho} \coloneqq   k \left( \ml{\rho} \mr{\rho}^{-1} \right) \mr{\rho} 
\end{align}
where $k: \mathbb R_{\ge 0} \to \mathbb R$ is an operator monotone function that satisfies
\begin{align}\label{eq:condition_k}
k(t) = t k\left( t^{-1} \right) \quad \forall t>0\,\;. 
\end{align}
\end{subequations}

It might be instructive at this point to check that the bilinear form $g_\rho$ indeed defines a Riemannian metric. Operator monotone functions are always analytic~\cite{bhatia2013matrix}, hence the bilinear form ~\eqref{eq:metric_k} is smooth with respect to $\rho$. Equation~\eqref{eq:condition_k} guarantees that $\mathcal K_\rho ^{-1}$ preserves hermiticity, making $g_{\rho}$ real valued. Since $\rho$ is a full-rank state, $\mathcal K_\rho ^{-1}$ is always well-defined and positive definite, implying that the bilinear form~\eqref{eq:metric_k} is non-degenerate, and therefore $g_{\rho}$ is indeed a Riemannian metric.

One can easily express $g_\rho (X,X)$ explicitly by invoking the spectral decomposition $\rho = \sum_i p_i \ket{i}\!\bra{i}$, which yields
\begin{align}
g_\rho (X,X) = \sum_{ij} \left[ p_i k\left( \frac{p_j}{p_i} \right) \right]^{-1} \left| \braket{i | X | j} \right|^2 \,\;.
\end{align}
Notice that, for all valid $k$, the diagonal part $i = j$ corresponds to the Fisher metric (up to proportionality constant). Consequently one recovers the  classical analogue result~\cite{campbell1986extended} over the $(d-1)$-dimensional probability simplex stating that if a Riemannian metric $g_{p}$ satisfies the monotonicity property
\begin{align}
g_{p} ( x,  x) \ge g_{T  (p)} (T (x), T (x))
\end{align}
under the action of stochastic matrices $T$ then it is necessarily proportional to the Fisher metric, i.e.,
\begin{align}
\left( g_p \right)_{ij} \propto \frac{\delta_{ij}}{p_i} \,\;.
\end{align}
A consequence of the richness of the monotone metrics in the quantum regime is the presence of the free parameter $\lambda$ in the monotones $\cq$ introduced in the main text.

Let us finally recall how inequality~\eqref{eq:main_inequality} implies the monotonicity property Eq.~\eqref{eq:metrics_data_processing}. In~\cite{lesniewski1999monotone}, the authors showed that for the class of functions $k$ defined above satisfying $k(1) = 1$ (which amount to a mere normalization condition), superoperators $\mathcal K_{\rho}^{-1} $ admit the  integral representation
\begin{align} \label{eq:integral_representation}
\mathcal K_{\rho}^{-1} = \int_0 ^\infty \left( \left[ \lambda \mr{\rho} + \ml{\rho}    \right]^{-1} + \left[  \mr{\rho} + \lambda \ml{\rho}    \right]^{-1}  \right) N_g(\lambda) d\lambda
\end{align}
where $N_g(\lambda) d\lambda$ is a singular measure (the detailed definition of can be found in~\cite{lesniewski1999monotone,temme2010chi}). The important point is that, due to the above representation, the monotonicity of the metric~\eqref{eq:metrics_data_processing} follows from the monotonicity of the integrand in Eq.~\eqref{eq:integral_representation}; this is what is shown in the theorem of Eq.~\eqref{eq:main_inequality}. In fact, notice that the theorem is slightly more general; it applies to any operator (i.e., not necessarily hermitian and traceless), a freedom that we take advantage of in the main text.

\end{document}